\documentclass[10pt, onecolumn]{article}

\usepackage{bm,color,bbm}
\usepackage{ulem} %for strike-through font, e.g \sout{Bush} Obama is the president.
\usepackage{hyperref, mathtools,graphicx} %COMMENTED OUT BY MR!!!
\usepackage{mathtools,graphicx}

\newcommand{\beq}{\begin{equation}}
\newcommand{\eeq}{\end{equation}}
\newcommand{\bqa}{\begin{eqnarray}}
\newcommand{\eqa}{\end{eqnarray}}

\newcommand{\id}{\mathbbm{1}}
\newcommand{\h}{{\cal H}}

\newcommand{\blk}{\color{black}}

\definecolor{maroon}{rgb}{0.7,0,0}

\definecolor{ngreen}{rgb}{0.3,0.7,0.3}

\definecolor{golden}{rgb}{0.8,0.6,0.1}

\begin{document}
\title{Entanglement of quantum fields\\
 via classical gravity}
\author{
	Marcel Reginatto\thanks{Physikalisch-Technische Bundesanstalt, Bundesallee 100, 38116 Braunschweig, Germany} ~ and Michael J. W. Hall\thanks{Department of Theoretical Physics, Research School of Physics and Engineering,
		Australian National University, Canberra ACT 0200, Australia}}
\date{\today}
\maketitle

\begin{abstract} We consider the coupling of quantum fields to classical gravity in the formalism of ensembles on configuration space, a model that allows a consistent formulation of interacting classical and quantum systems. 
 Explicit calculations show that there are solutions for which two quantum fields are in an entangled state, even though their interaction occurs solely via a common classical gravitational field, and that such entangled solutions can evolve from initially unentangled ones.
These results support the observation of a previous paper that an observed generation of entanglement would not provide a definitive test of the nonclassical nature of gravity.
\end{abstract}

\section {Introduction}\label{intro}

The publication of recent proposals for witnessing nonclassical features of gravity, by Bose {\it et al.} \cite{bose} and by Marletto and Vedral \cite{marl},  has encouraged a discussion about the possibility of generating entanglement between quantum systems which only interact via a classical gravitational field, as well as some new proposals for looking for evidence of quantum gravity in laboratory experiments \cite{HR2018,MV201803,MV201804,AH2018,BWGCBA2018,CR2018}. Whether entanglement is possible under these circumstances depends on which hybrid model is used to describe the interaction of classical and quantum sectors \cite{HR2018}. While some hybrid models of classical-quantum interactions seem to exclude entanglement (e.g., Koopman-type dynamics and mean-field models), other models, in particular the formalism of ensembles on configuration space \cite{hrbook}, allow for it. Thus an observed generation of entanglement cannot provide a definitive test of the nonclassicality of gravity without additional assumptions concerning the nature of classical-quantum interactions.

In our previous paper \cite{HR2018}, we provided simplified examples where entanglement generation between two quantum systems, via a classical mediating system, was possible even though there was no direct interaction between the quantum systems. 
We also argued that similar conclusions could be expected for gravitational models of classical-quantum interactions, but we did not provide any explicit calculations for the case in which the classical sector was a gravitational field. The purpose of this paper 
%is to provide a follow up to the previous one and
is to present fully relativistic calculations showing that (i) there are indeed solutions for which two quantum fields are in an entangled state even though their interaction occurs solely via a common classical gravitational field, and (ii) such entangled solutions can evolve from initially unentangled ones. 

While a fully relativistic calculation is obviously more complicated than one based on the weak-gravity, non-relativistic limit, it is desirable so that the dynamical degrees of freedom of the gravitational field are taken into consideration (on this point, see remarks in Refs. \cite{AH2018} and \cite{BWGCBA2018}). Another reason for carrying out a detailed relativistic calculation within a particular hybrid model is to examine the limitations of such models: since the quantization of gravity does not appear to follow from consistency arguments alone \cite{C11_AKR2008}, it is of interest to investigate to what extent hybrid systems can provide a satisfactory description of matter and gravitation. The study of such systems may provide valuable clues that can be of help in the search for a full quantum theory of gravity.

The paper is organized as follows. In section \ref{ECS}, we give a brief introduction to the basic aspects of the configuration ensemble approach using the example of non-relativistic particles. In section \ref{QFCG}, we discuss the extension to quantum matter fields coupled to a classical spacetime described by the Einstein equations, focusing on the case where the quantum sector consists of two quantized scalar fields.  We also consider the particular case of spherical gravity, a midisuperspace formulation applicable to the case of spherical symmetry. In the following two sections we provide approximate solutions for spherical gravity. In section \ref{EntQFBH}, we consider a perturbative approach and show that there are solutions for which the two quantum fields are in an entangled state. In section \ref{EoT}, after showing that one may introduce a gravitational time, we discuss how entangled solutions can evolve from initially unentangled ones. Finally, in section \ref{Discussion}, we provide some concluding remarks and discuss possible future directions.

\section{Ensembles on configuration space describing classical,
quantum and mixed classical-quantum systems}\label{ECS}

This section and the one that follows present those aspects of the formalism of ensembles on configuration space that are needed for the calculation of entanglement presented in section \ref{EntQFBH}. As the approach is not well known, we have tried to provide a fairly complete summary.

The formalism of ensembles on configuration is a general framework that allows for the description of classical, quantum and hybrid systems \cite{hrbook,hr2005,consist}. It is underpinned by a very simple physical picture: ensembles evolving on a configuration space. The mathematical structure is correspondingly simple (simpler than $C^*$-algebras for example), yet is sufficiently nontrivial to guarantee the existence of, for example, a  dynamical  bracket for observables, thermal ensembles, weak values, and a generalised Ehrenfest theorem. It also forms a natural starting platform for several axiomatic approaches to reconstructing quantum theory \cite{hrbook,C1_HR2002,C1_HKR2003,C1_RH2011,C1_RH2012,C1_R2014}.  Further, such models have been applied to model gravity via the coupling of ensembles of quantum fields to classical spacetimes \cite{hrbook,hr2005,ark,marcel}.

\subsection{Classical, quantum, and mixed classical-quantum systems}

We start from the assumption that the configuration of a physical system is an inherently statistical concept. The system will therefore be described by an ensemble of configurations, with probability density $P$, with $P \geq 0$ and $\int dx \,P(x,t) = 1 $ (the case in which the uncertainty goes to zero can be described in the obvious way, by setting $P$ equal to a delta function, so systems with negligible uncertainty are also included in the formalism).

To describe dynamics, we introduce an {\it ensemble Hamiltonian}
${H}[P,S]$, where $S$ is an auxiliary field that is canonically conjugate to $P$. The equations of motion take the form 
\begin{equation}\label{PoissonBrackets}
\frac{\partial P}{\partial t} = 
%\left\{ P,{H} \right\}_{PB}
 \frac{\delta{H}}{\delta S},~~~~~~~~~\frac{\partial S}{\partial t} = 
 %\left\{ S,{H} \right\}_{PB}
-\frac{\delta{H}}{\delta P},
\end{equation}
where 
$\delta/\delta P$ ($\delta/\delta S)$) denotes the functional derivative with respect to $P$ ($S$). 

The following ensemble Hamiltonians lead to equations that describe
the evolution of quantum and classical non-relativistic  particles of mass $m$: 
\begin{eqnarray}
{H}_C[P,S] &=& \int dx\, P \left[ \frac{|\nabla S|^2}{2m} + V(x)\right] ,\label{HC_App}\\
{H}_Q[P,S] &=& {H}_C[P,S]
+  \frac{\hbar^2}{4} \int dx\ P\frac{|\nabla \log P|^2}{2m} .\label{HQ_App}
\end{eqnarray}
For example, the equations of motion derived from ${H}_Q[P,S]$ are given by
\begin{equation}\label{QEqMotion_App}
\frac{\partial P}{\partial t} + \nabla .\left( P\frac{\nabla S}{m} \right) =0,~~~~~~~\frac{\partial S}{\partial t} + \frac{|\nabla S|^2}{2m} + V -  \frac{\hbar^2}{2m}\frac{\nabla^2 P^{1/2}}{P^{1/2}} = 0
\end{equation}
while the equations of motion derived from ${H}_C[P,S]$ are
the same as Eq. (\ref{QEqMotion_App}) but with $\hbar=0$. The first
equation in Eq. (\ref{QEqMotion_App}) is a continuity equation, the
second equation is the classical Hamilton-Jacobi equation when
$\hbar = 0$ and a modified Hamilton-Jacobi equation when $\hbar \neq
0$. Defining $\psi=\sqrt{P}~e^{iS/\hbar}$, Eq.~(\ref{QEqMotion_App}) takes the form
\begin{equation}
i\hbar \frac{\partial \psi}{\partial t}
= \frac{-\hbar^2}{2m}\nabla^2\psi + V\psi,
\end{equation}
which is the usual form of the Schr\"{o}dinger equation. 

It is straightforward to extend the formalism in a natural way to allow for mixed quantum-classical systems. A mixed quantum-classical ensemble
Hamiltonian on a joint configuration space with coordinates $q$, $x$ is given by
\begin{eqnarray}\label{HQC_App}
{H}_{QC}[P,S] &=& \int dq\,dx\, P\,\left[ \frac{|\nabla_x S|^2}{2M}
+ \frac{|\nabla_q S|^2}{2m}
 \right] \nonumber\\
&+& \int dq\,dx\, P\,\left[  \frac{\hbar^2}{4} \frac{|\nabla_q \log P|^2}{2m} + V(q,x,t)\right] .
\end{eqnarray}
 Here $q$ denotes the configuration space coordinate of a quantum
particle of mass $m$ and $x$ that of a classical particle of mass
$M$, and $V(q,x,t)$ is a potential energy function describing the
quantum-classical interaction. The equations of motion for the joint probability density $P(q,x)$ and its conjugate 
$S(q,x)$ follow from ${H}_{QC}$ as 
\begin{eqnarray}\label{EqsCQ_App}
\frac{\partial P}{\partial t} &=&  -\nabla_q .\left( P
\frac{\nabla_q S}{m} \right) - \nabla_x .\left(P\frac{\nabla_x S}{M}\right), \nonumber\\
\frac{\partial S}{\partial t} &=& -
\frac{|\nabla_q S|^2}{2m} - \frac{|\nabla_x S|^2}{2M} - V +
\frac{\hbar^2}{2m}\frac{\nabla_q^2 P^{1/2}}{P^{1/2}} .
\end{eqnarray}
These can be rewritten as a nonlinear Schr\"odinger equation for the `hybrid' wave function $\psi=\sqrt{P}e^{iS/\hbar}$, and one obtains a similar nonlinear equation for the case of two fully classical particles. Such nonlinearity does not automatically lead to difficulties in either case, essentially because the form of classical observables is fundamentally different to that of quantum observables, as discussed below. 

\subsection{Observables}\label{subsecobs}

The \textit{state} of a system is completely determined by specifying the two fields $P$ and $S$. \textit{Observables} are defined as suitable functionals of $P$ and $S$. Given two functionals $A[P,S]$ and $B[P,S]$, define their Poisson bracket in the standard way,
\begin{equation}\label{pb}
\{ A,B \}_{PB} = \int dx \left( \frac{\delta A}{\delta P} \frac{\delta B}{\delta S} - \frac{\delta A}{\delta S} \frac{\delta B}{\delta P} \right).
\end{equation}
This gives us an \textit{algebra of obervables}.

Arbitrary functionals $A[P,S]$ are not necessarily observables because these have to satisfy certain mild requirements. For example, the infinitesimal canonical transformation generated by any observable $A$ must preserve the normalization and positivity of
$P$. This implies the two conditions
\begin{equation}
A[P,S+c] = A[P,S],~~~~~\delta A / \delta S = 0 ~ \textrm{if} ~
P(x)=0.
\end{equation}
Note that the first equation implies gauge invariance of the theory under $S \rightarrow S + c$. A more general condition that may be imposed on observables is that they be homogeneous of degree one in $P$; i.e., $A[\lambda
P,S]=\lambda A[P,S]$. Then, it follows that
\begin{equation}
A[P,S] = \int dx \,P \left( \delta A / \delta P \right) = \langle \delta A/ \delta P \rangle .
\end{equation}
That is, one can associate with each observable $A$ a local density on the configuration space, and the value of $A$ can be calculated by integrating over this local density.

There is a critical physical distinction between classical and quantum systems (or classical and quantum components of a composite hybrid system): they have quite different sets of observables, and distinct algebras for these observables \cite{hrbook}.

For example, for a purely classical configuration space labelled by position $x$, the classical observable $C_f$ corresponding to the phase space function $f(x,k)$ (where $k$ is the momentum) is defined by the functional
\beq \label{cf}
C_f[P,S] := \int dx\, P\,f(x,\nabla_x S) .
\eeq
Note that it is numerically equal to the ensemble average of $f(x,k)$, provided one associates momentum $k=\nabla_x S$ with position $x$.  Evaluating the Poisson bracket of any two classical observables $C_f, C_g$ via Eq.~(\ref{pb}) yields
\beq \label{cb}
\{C_f,C_g\} = C_{\{f,g\}},
\eeq
where $\{f,g\}=\sum_i \left(\frac{\partial f}{\partial x_i}\frac{\partial g}{\partial k_i} - \frac{\partial f}{\partial k_i}\frac{\partial g}{\partial x_i}\right)$ denotes the usual phase space bracket. Thus, \textit{the Poisson bracket for classical ensembles is isomorphic to the classical phase space bracket}, implying that it generates the standard classical dynamics (and a continuity equation for $P$~\cite{hrbook,consist}).  Eq.~(\ref{cf}) is an  isomorphism  between  the algebra of observables $C_f$ on configuration space  and  the algebra of observables $f$ on classical phase space.

Similarly, for a purely quantum configuration space labelled by the possible outcomes $q$ of some complete basis set $\{|q\rangle\}$ of a Hilbert space $\cal H$ (i.e., $\int dq\,|q\rangle\langle q|=\hat \id$, with integration replaced by summation for discrete ranges of $q$), the quantum observable $Q_{\hat M}$ corresponding to the Hermitian operator $\hat M$ is defined by the functional
\beq \label{qm}
Q_{\hat M}[P,S] := \langle \psi|\hat M|\psi\rangle,
\eeq
where $|\psi\rangle\in \h$ is the wave function defined via $\langle q|\psi\rangle=\sqrt{P(q)}e^{iS(q)/\hbar}$. Note that $Q_{\hat M}$ is numerically equal to the ensemble average of $\hat M$ for quantum state $|\psi\rangle$. Further, evaluating the Poisson bracket of any two quantum observables $Q_{\hat M}, Q_{\hat N}$ via Eq.~(\ref{pb}) yields~\cite{hrbook,consist}
\beq \label{qb}
\{ Q_{\hat M}, Q_{\hat N}\} = Q_{[\hat M,\hat N]/(i\hbar)},
\eeq
where $[\hat M,\hat N]$ is the usual commutator. Thus, \textit{the Poisson bracket for quantum ensembles is isomorphic to the quantum commutator}, implying that it generates the usual Schr\"odinger equation.  Eq. (\ref{qm})  is an isomorphism between the algebra of observables $Q_{\hat M}$ on configuration space and the algebra of quantum observables $\hat M$.

The Poisson bracket properties (\ref{cb}) and (\ref{qb}) remain unchanged in the case of mixed classical-quantum systems of the type described by the ensemble Hamiltonian of Eq. (\ref{HQC_App}). Thus, since the bracket is preserved under Hamiltonian evolution, the algebra of classical observables remains isomorphic to the classical phase space bracket, and the algebra of quantum observables remains isomorphic to the quantum commutator -- even under interactions between the classical and quantum components. In particular, the classical/quantum distinction is always maintained. 

Thus there are two important ways in which the classical or quantum nature of a subsystem that belongs to a composite system is manifested:
\begin{enumerate}
\item By the form of the ensemble Hamiltonian when restricted to the classical and quantum sectors, which is of the form of Eq. (\ref{HC_App}) for the classical case and of Eq. (\ref{HQ_App}) for the quantum case, and
\item By the fact that the algebra of classical observables remains isomorphic to the classical phase space bracket, and the algebra of quantum observables remains isomorphic to the quantum commutator.
\end{enumerate}

\subsection{Entanglement}\label{entanglement}

Two ensembles with respective configuration spaces $X$ and $Y$ are defined to be independent if $P(x,y)$ and $S(x,y)$ satisfy~\cite{hrbook, hr2005}
\begin{equation}
P(x,y) = P_X(x)P_Y(y),~~~~~S(x,y)=S_X(x)+S_Y(y) 
\end{equation}
(with the latter only required to hold up to some additive constant, recalling invariance under $S\rightarrow S+c$). For quantum ensembles, note that independence is equivalent to a factorisable wave function $\psi=\sqrt{P}e^{iS/\hbar}$, and hence any two quantum ensembles are either independent or entangled. But 
can entanglement be defined for more general physical systems that are described using the formulation of ensembles on configuration space? 

As it turns out, the concept of entanglement remains meaningful in the general case \cite{hrbook}. However, it is important to note that the notion of `entanglement' referred to here is not in the strong sense of Bell inequality violation, but in Schr\"odinger's original weaker sense that the properties of a joint ensemble cannot be decomposed into properties of the individual ensembles \cite{C3_S1935} (Spekkens has similarly used this weaker sense to define entanglement for a class of `epistemic' models of statistical correlation \cite{C3_S2007}).

This can be understood by looking at a simple classical example. In particular, consider a classical joint ensemble, corresponding to two classical particles described by respective configuration spaces $X$ and $Y$, with probability density $P(x,y)$ and conjugate quantity $S(x,y)$. Noting that the product of two classical phase space functions $f(x,p_x)$ and $g(y,p_y)$ is itself a classical phase space function for the two particles, we recall from section~\ref{subsecobs} that the expectation value of this product  corresponds to the classical observable
\beq \label{cfg}
C_{fg} = \langle fg\rangle = \int d x d y\, P(x,y)\,f(x,\partial_x S)\,g(y,\partial_y S).
\eeq
Now, there is clearly a trivial hidden variable for any such observable. In particular, defining $\lambda:=[x,y,S(x,y)]$, $P(\lambda):=P(x,y)$,  $F(\lambda):= f(x,\partial_x S)$, and $G(\lambda):=g(y,\partial_y S)$, one has
\beq
\langle fg\rangle = \int d \lambda\, P(\lambda)\, F(\lambda)\,G(\lambda).
\eeq
Hence, no Bell inequality can be violated via such observables \cite{C3_HHHH2009}.

Nevertheless, if the independence condition $S(x,y)=S_X(x)+S_Y(y)$ is \textit{not} satisfied, then the `hidden value' of the observable $f(x,p_x)$ for the {\it first} particle, i.e., $F(\lambda)$, will in general depend on the position of the {\it second} particle, via $p_x=\partial_x S(x,y)$.  That is, while knowledge of the position and momentum of the first particle at a given time is sufficient to determine all observables for the particle at that time, it will not be sufficient to determine them at any later time: one needs to know the evolution of the joint quantity $\partial _x S(x,y)$.  Moreover, if one locally perturbs the position of the second particle, from $y$ to $y'$, the corresponding perturbation of $S(x,y)$ to $S(x,y')$ will typically perturb the value of $p_x$ in this model.

Hence, \textit{a kind of nonlocality, or inseparability, can be associated even with classical configuration space ensembles}. We will, by analogy with Schr\"odinger's original discussion \cite{C3_S1935}, refer to this property as `entanglement'. This leads to the following general definition which applies to {\it all} configuration space ensembles \cite{hrbook}:
	\begin{quote} \it
		A joint ensemble is entangled with respect to the joint configuration space $X\times Y$  if and only if $S(x,y)\not\equiv S_X(x)+S_Y(y)$ (up to some additive constant).
	\end{quote}
%}
Note that entanglement, as defined here, is relative to  particular configuration spaces $X$ and $Y$. For quantum ensembles, this corresponds to a particular choice of computational basis for the component ensembles.  For this reason, our definition of entanglement is stronger than the standard definition for quantum ensembles (where the latter requires only that $S(x,y)\neq S_X(x)+S_Y(y)$ for {\it some} choice of computational basis, i.e, that the ensembles are not independent). We require a stronger definition because for general configuration ensembles one does not have a similar freedom to arbitrarily choose between configuration spaces.

Note also that entanglement as defined above is not the converse of  independence, in contrast to the quantum case. Indeed, it is shown in section~3.2.3 of \cite{hrbook} that it is natural to regard two classical systems as unentangled even if $P(x,y)\neq P(x)P(y)$, providing that $S(x,y)=S_X(x)+S_Y(y)$. Finally, we remark that the above definition can be extended in an obvious manner to more than two systems.

\section{The coupling of scalar quantum fields to classical gravity}\label{QFCG}

We now consider the coupling of scalar quantum fields to classical gravity using configuration space ensembles, and write down the relevant equations. A detailed description of the formalism is given in~\cite{hrbook}.

\subsection{General case}

We proceed by steps: we first define a classical configuration space ensemble for pure gravity and then consider the addition of quantum fields.

\subsubsection{Vacuum gravity}
\label{vacgrav}

The most direct way of introducing a classical configuration space ensemble for vacuum gravity is to start from the Einstein-Hamilton-Jacobi equation, which in the metric representation takes the form \cite{MTW73,K12}
\begin{equation}\label{Appendix-ehj}
H^C_h =\frac{16\pi G}{c^2} G_{ijkl}\frac{\delta S}{\delta h_{ij}}\frac{\delta S}{
\delta h_{kl}}-\frac{c^4}{16\pi G }\sqrt{h}\left( R-2\lambda \right) =0 .
\end{equation}
Here $G$ is the gravitational constant; $c$ is the speed of light; $h_{ij}$ is the spatial part of the metric tensor, with determinant $h$; $R$\ is the three-dimensional curvature scalar; $G_{ijkl}=\frac{1}{2\sqrt{h}}\left( h_{ik}h_{jl}+h_{il}h_{jk}-h_{ij}h_{kl}\right) $ is the
DeWitt supermetric \cite{MTW73}; and $\lambda $ is the cosmological constant.

We will assume that the functional $S$ is invariant under the gauge group of spatial coordinate transformations, which is equivalent to satisfying the momentum (or diffeomorphism) constraints of the canonical formulation of general relativity.

There is no need to fix a particular gauge when solving the Einstein-Hamilton-Jacobi equation:  Equation~(\ref{Appendix-ehj}) is valid for an arbitrary choice of lapse function $N$ and foliation and corresponds to an infinity of constraints, one at each point. However, one may introduce an alternative viewpoint \cite{K93}, where equation~(\ref{Appendix-ehj}) is regarded as an equation to be integrated with respect to a ``test function'' in which case we are dealing with one equation for each choice of lapse function $N$,
\begin{equation} \label{Appendix-hg}
\int d^{3}x\,\,N H^C_h=0;
\end{equation}
i.e., for each choice of foliation. Such an alternative viewpoint is extremely useful: although it may be impossible to find the general solution (which requires solving the Einstein-Hamilton-Jacobi equation for all choices of lapse functions), it may be possible to find particular solutions for specific choices \cite{K93}.

Taking Eq. (\ref{Appendix-hg}) into consideration, we define the {\it ensemble} Hamiltonian for vacuum gravity according to
\begin{equation} \label{Appendix-eehj}
{\cal H}^C_h=\int d^{3}x N \int Dh\,P\,\,H^C_h,
\end{equation}
where $P$ is a probability density function and $Dh$ is an appropriate measure over the space of metrics (technical issues are discussed in~\cite{hrbook}). The functional $P$ is also assumed, like $S$, to be invariant under the gauge group of spatial coordinate transformations. The corresponding equations of motion have the form
\begin{equation}
\frac{\partial P}{\partial t}=\frac{\Delta {\cal H}^C_h}{\Delta S},\quad
\frac{\partial S}{\partial t}=-\frac{\Delta {\cal H}^C_h}{\Delta P}, 
\end{equation}
where $\Delta /\Delta F$ denotes the variational derivative with respect to a functional $F$ (see Appendix A of~\cite{hrbook}). Assuming the constraints $\frac{\partial S}{\partial t}=$ $\frac{\partial P}{\partial t}=0$, these equations lead to equation~(\ref{Appendix-hg}), as required, and to a continuity equation,
\begin{equation} \label{Appendix-cehj0}
\int d^{3}x \,N \frac{\delta }{\delta h_{ij}}\left( P\,G_{ijkl}\frac{\delta S}{
\delta h_{kl}}\right) =0.
\end{equation}
These two equations, Eqs. (\ref{Appendix-hg}) and (\ref{Appendix-cehj0}), define the evolution of an ensemble of classical spacetimes on configuration space for the case of vacuum gravity.

\subsubsection{The addition of quantum scalar fields}

A hybrid system, where $n$ quantum scalar fields $\phi_a$ of mass $m_a$ independently couple to the classical metric $h_{kl}$, requires a generalization of equation~(\ref{Appendix-eehj}) in which
\begin{equation} \label{Appendix-ehqphih}
{\cal H}_{\phi h}=\int d^{3}x \,N \int Dh D\phi \,P  \left[ {H}^C_{\phi h}+F_{\phi }
\right] ,
\end{equation}
where
\begin{equation}
{H}^C_{\phi h}= {H}^C_{h}+
\sum_{a=1}^n \left\{ \frac{1}{2\sqrt{h}}\left( \frac{\delta S
}{\delta \phi_a }\right) ^{2}+\sqrt{h}\left[ \frac{1}{2}h^{ij}\frac{\partial
\phi_a }{\partial x^{i}}\frac{\partial \phi_a }{\partial x^{j}}+ \frac{1}{2}m_a^2\phi_a
^2 \right] \right\} \nonumber
\end{equation}
and
\begin{equation}
F_{\phi }=\sum_{a=1}^n \left\{ \frac{\hbar ^{2}}{4}\frac{1}{2\sqrt{h}}\left( \frac{\delta \log P
}{\delta \phi_a }\right) ^{2} \right\}. \nonumber
\end{equation}
To interpret the terms that appear in Eq. (\ref{Appendix-ehqphih}), note that ${H}^C_{\phi h}=0$ is the Einstein-Hamilton-Jacobi equation for gravity with $n$ {\it classical} scalar fields and that $F_{\phi }$ is a non-classical term which accounts for the quantum nature of the fields.

Assuming again the constraints $\frac{\partial S}{\partial t}=$ $\frac{\partial P}{\partial t}=0$, the corresponding equations are given by
\begin{equation} \label{Appendix-hh}
\int d^{3}x\, N \, \left[ \mathcal{H}^C_{\phi h}
-\sum_{a=1}^n \left\{ \frac{\hbar ^{2}}{2\sqrt{h}}
\left( \frac{1}{\sqrt{P}}\frac{\delta ^{2} \sqrt{P}}{\delta \phi_a ^{2}}\right) \right\} \right]=0,
\end{equation}
and a continuity equation of the form
\begin{equation} \label{Appendix-cehj}
\int d^{3}x \,N \left[ \frac{32\pi G}{c^2} \frac{\delta }{\delta h_{ij}}\left( P\,G_{ijkl}\frac{\delta S}{
\delta h_{kl}}\right) +  \sum_{a=1}^n \left\{ \frac{1}{\sqrt{h}} \frac{\delta }{\delta \phi_{a}}\left( P \frac{\delta S}{
\delta \phi_{a}}\right) \right\} \right] =0.
\end{equation}
These are the equations that need to be solved in the general case.

\subsection{Spherical symmetry}

In the case of spherical symmetry, it is possible to give a midisuperspace formulation of general relativity known as spherical gravity. It leads to a simpler set of equations.

For spherical symmetry, the line element may be written in the form
\begin{equation} \label{1022_01}
g_{\mu \nu }dx^{\mu }dx^{\nu }=-N^{2}dt^{2}+\Lambda ^{2}\left(dr+N_{r}dt\right) ^{2}+R^{2}d\Omega ^{2}.
\end{equation}
The lapse function $N$ and the shift function $N_r$ are functions of the radial coordinate $r$ and the time coordinate $t$. The configuration space for the gravitational field consists of two fields, $R$ and $\Lambda$. Spherically symmetric gravity is discussed in detail in a number of papers, mostly in reference to the canonical quantization of black hole spacetimes. For discussions using the metric representation, see for example \cite{C10_R95, C10_K94, C10_L95}. For discussions of the Einstein-Hamilton-Jacobi equation in the context of the WKB approximation of quantized spherically symmetric gravity, see for example \cite{C10_FMP90, C10_BK97}.

\subsubsection{Vacuum gravity}

To simplify the equations, we set $c=G=\hbar=1$ %\michael{Should be $16\pi g$?}\marcel{Hmmm... I need to check that.} \michael{I'm just going from Eq. 17.}\marcel{It turns out that $G=1$ is correct after all. The procedure that takes you to spherical gravity is worked out in detail in: Kuchar, ``Geometrodynamics of Schwarschild black holes,'' Phys. Rev. D 50, 3961-3981 (1994), sections IIIA and IIIB. To go from the four-dimensional space-time with coordinates $(t,r,\theta,\phi)$ to the two-dimensional space-time with coordinates $(t,r)$ you need to integrate over angles $\theta$ and $\phi$ and you pick up a factor of $4\pi$. The additional factor of 4 comes from expressing the curvature scalar and the extrinsic curvature in terms of the line element of Eq. (\ref{1022_01}). Amazing but true.} 
from now on. The Einstein-Hamilton-Jacobi equation for the case of vacuum gravity takes the form $H_{\Lambda R}=0$ with
\begin{equation} \label{1022_02}
H_{\Lambda R} = -\frac{1}{R}\frac{\delta S}{\delta R}\frac{\delta S}{\delta
	\Lambda }+\frac{\Lambda }{2R^{2}}\left( \frac{\delta S}{\delta \Lambda }
\right) ^{2} + V,
\end{equation}
where
\begin{equation} \label{1022_02a}
V=\frac{RR^{\prime \prime }}{\Lambda }-\frac{RR^{\prime }\Lambda ^{\prime }}{\Lambda ^{2}}
+\frac{R^{\prime 2}}{2\Lambda }-\frac{\Lambda }{2}.
\end{equation}
In spherical gravity, the three momentum constraints of the full theory are replaced by a single (radial) diffeomorphism constraint,
\begin{equation} \label{1022_03}
\frac{\delta S}{\delta R}R^{\prime }-\Lambda \left( \frac{
	\delta S}{\delta \Lambda }\right) ^{\prime }=0,
\end{equation}
where primes indicate derivatives with respect to $r$ \cite{K12}. One can see from this constraint that $R$ transforms as a scalar while $\Lambda$ transforms as a scalar density. 
We will require that $S$ be invariant under diffeomorphisms so that it automatically solves the momentum constraint. 

An appropriate
ensemble Hamiltonian for spherically symmetric gravity is given by 
\begin{equation} \label{1022-03a}
{\cal H} =\int d r \,N \int \mathrm{D}R \, \mathrm{D} \Lambda 
\,P H_{\Lambda R}.
\end{equation}

With $\frac{\partial S}{\partial t}=\frac{\partial P}{\partial t}=0$ again, and assuming $N$ is arbitrary, the equations of motion derived from the ensemble Hamiltonian of Eq. (\ref{1022-03a}) are Eq. (\ref{1022_02}), the Einstein-Hamilton-Jacobi equation,
\begin{equation} \label{1022-03c}
H_{\Lambda R} = -\frac{1}{R}\frac{\delta S}{\delta R}\frac{\delta S}{\delta
	\Lambda }+\frac{\Lambda }{2R^{2}}\left( \frac{\delta S}{\delta \Lambda }
\right) ^{2} + V=0,
\end{equation}
and the continuity equation
\begin{equation}\label{1022_04}
\frac{\delta }{\delta R}\left( P\frac{1}{R}\frac{\delta S}{\delta \Lambda }\right) + \frac{\delta }{\delta \Lambda }\left( P\frac{1}{R}\frac{\delta S}{\delta R}- P \frac{\Lambda }{R^{2}}\frac{\delta S}{\delta \Lambda }\right) =0.
\end{equation}

\subsubsection{The addition of quantum scalar fields}

The ensemble Hamiltonian of a hybrid system where matter is in the form of $n$ minimally coupled quantized radially symmetric scalar fields $\phi_a$ of mass $m$ is given by
\begin{equation} \label{1142_01}
{\cal H}_{\phi \Lambda R}=\int d r\int \mathrm{D} \phi \mathrm{D} \Lambda \mathrm{D} R \;  P\,N \left[ {H}^C_{\phi \Lambda R}+F_{\phi } \right] ,
\end{equation}
where
\begin{equation}\label{1142_02}
{H}^C_{\phi \Lambda R}={H}^C_{\Lambda R}
+\sum_{a=1}^n \left[\frac{1}{2\Lambda R^{2}}\left( \frac{\delta S}{\delta \phi_a }\right) ^{2}
+\frac{R^2 }{2 \Lambda}\phi_a ^{\prime 2} + \frac{\Lambda R^{2} m_a^2}{2 } \phi_a ^{ 2} \right],
\end{equation}
is a purely classical term which now includes the coupling to scalar field $\phi_a$ and
\begin{equation}\label{1142_03}
F_{\phi }= \frac{1}{8 \Lambda R^2} \sum_{a=1}^n \left( \frac{\delta \log P
}{\delta \phi_a }\right) ^{2}
\end{equation}
is an additional, non-classical term that must be included in the ensemble Hamiltonian when the scalar field is quantized  (recall we have set $\hbar=1$). 

Assuming again the constraints $\frac{\partial S}{\partial t}=$ $\frac{\partial P}{\partial t}=0$, the corresponding equations are
\begin{equation} \label{1142_04}
\int dr\, N \, \left[ {H}^C_{\phi \Lambda R}
-\frac{1}{2\Lambda R^2} \sum_{a=1}^n
\left( \frac{1}{\sqrt{P}}\frac{\delta ^{2}\sqrt{P}}{\delta \phi_a ^{2}}\right) \right]=0,
\end{equation}
and the continuity equation
\begin{eqnarray} \label{1142_05}
\int dr\, N \, \left[
\frac{\delta }{\delta R}\left( P\frac{1}{R}\frac{\delta S}{\delta \Lambda }\right)
+\frac{\delta }{\delta \Lambda }\left( P\frac{1}{R}\frac{\delta S}{\delta R}
-P\frac{\Lambda }{R^{2}}\frac{\delta S}{\delta\Lambda }\right)\right.&~&\nonumber\\
\left.- \sum_{a=1}^n  \frac{\delta }{\delta \phi_a }\left( P\frac{1}{\Lambda R^{2}}
\frac{\delta S}{\delta \phi_a }\right) \right]&=&0.
\end{eqnarray}

\section{Black hole with two scalar quantum fields in spherical gravity: entangled solutions}\label{EntQFBH}

We want to consider the following example: two quantized scalar fields, $\phi_1$ and $\phi_2$, in the space-time of a classical black hole, under the assumption that the quantum fields act as a perturbation to the space-time; i.e., that the contribution to the gravitational field from the mass of the black hole is much larger than that of the quantum matter fields. We will work out a midisuperspace solution in spherical gravity.

Under these circumstances, it is appropriate to search for an approximate perturbative solution of Eqs. (\ref{1142_04}) and (\ref{1142_05}) based on an expansion in powers of $\phi_a$ \cite{S97}.  The advantage of using such an approach is that it is possible to solve the equations iteratively, term by term, as is clear from the equations below. We will use the notation of \cite{S97} where $S^{(n)}$ stands for a functional of order $(\phi_a)^n$. While the term $S^{(0)}$ can be chosen freely, the higher order terms depend on the previous ones.

To carry out the calculation, it will be convenient to write the expression for $P[R,\Lambda,\phi_1,\phi_2]$ in the form
\begin{eqnarray}
P &=& e^{-\left(\sum_k F^{(k)}[R,\Lambda,\phi_1,\phi_2]\right)} = e^{-\left(\sum_n F^{(0)}[R,\Lambda]\right)} \, e^{-\left(\sum_{n>0} F^{(n)}[R,\Lambda,\phi_1,\phi_2]\right)}\nonumber\\
&=:& P_A[R,\Lambda] \, P_B[R,\Lambda,\phi_1,\phi_2]
\end{eqnarray}
so that $P_A$ depends only on the gravitational degrees of freedom while $P_B$ depends on both gravitational and scalar field degrees of freedom. Furthermore, we will require that, \textit{up to order $(\phi_a)^2$},
\begin{eqnarray}
\frac{\delta P_B}{\delta \phi_a } &=& -\frac{\delta F^{(2)}}{\delta \phi_a }P_B,\\
\frac{\delta P_B}{\delta h_{ij} } &=& -\frac{\delta F^{(2)}}{\delta h_{ij} }P_B, \\
\frac{1}{\sqrt{P_B}} \frac{\delta ^{2}\sqrt{P_B}}{\delta \phi_a^{2}} &=& -\frac{1}{2}\frac{\delta^2 F^{(2)}}{\delta \phi_a^2} + \frac{1}{4} \left(\frac{\delta F^{(2)}}{\delta \phi_a }\right)^2, 
\end{eqnarray}
(note the terms on the right of the last equation are of order $(\phi_a)^0$ and $(\phi_a)^2$) respectively). Our ansatz then is that the odd terms vanish, i.e., $F^{(1)}=F^{(3)}=0$, so that $P_B$ is to a first approximation a \textit{Gaussian functional}\footnote{While this is not an essential assumption, this choice for $P_B$ seems physically reasonable since it implies a solution of the quantum sector that is in some respect close to the simplest solution that one gets in quantum field theory in curved space time (i.e., the ground state functional).} with respect to the $\phi_a$. This approximation is sufficient for the approximation discussed below, which only considers solving $S^{(n)}$ to up to order $n=2$ . 

The expression ${\delta^2 F^{(2)}}/{\delta \phi_a^2 }$ needs to be regularized (such a term appears also in solutions of the Schr\"{o}dinger functional equation). We will not consider the regularization problem here, we will simply assume that this term has been regularized and it is finite. We will assume ${\delta^2 F^{(2)}}/{\delta \phi_1^2 }={\delta^2 F^{(2)}}/{\delta \phi_2^2 }$ and introduce the notation
\begin{equation}
\frac{1}{2}\frac{\delta^2 F^{(2)}}{\delta \phi_a^2} =: C_F[R,\Lambda]
\end{equation}
for this term.

We now give explicit solutions for $S^{(n)}$ that are valid to up to order $n=2$, and discuss the equation that determines the next term of the series expansion in powers of $\phi_a$, focusing on the question of the existence of solutions with entangled quantum fields: 
\begin{enumerate}
	\item Zeroth order terms:
	\begin{eqnarray}
&~&\left[ -\frac{1}{R}\frac{\delta S^{(0)}}{\delta R}\frac{\delta S^{(0)}}{\delta
	\Lambda } + \frac{\Lambda }{2R^{2}}\left( \frac{\delta S^{(0)}}{\delta \Lambda }
\right) ^{2} +  \frac{RR^{\prime \prime }}{\Lambda }-\frac{RR^{\prime }\Lambda ^{\prime }}{\Lambda ^{2}}
+\frac{R^{\prime 2}}{2\Lambda }-\frac{\Lambda }{2} \right]   \nonumber\\
	&~& + \sum_{a=1}^2 \left\{ \frac{1}{2\Lambda R^{2}}\left( \frac{\delta S^{(1)}}{\delta \phi_a }\right) ^{2} - \frac{C_F}{8\Lambda R^{2}} \right\} = 0.
	\end{eqnarray}
	We choose the $S^{(0)}$ that solves the classical EHJ equation for a black hole; i.e, that makes the terms in square brackets equal to zero. This solution is known \cite{C10_BK97}, 
	\begin{equation}\label{1030_05}
	S^{(0)} = \int dr\left\{ \Lambda R\sqrt{\frac{R^{\prime 2}}{\Lambda^{2}}+\frac{2m}{R}-1} -RR^{\prime }\cosh ^{-1}\left(\frac{R^{\prime }}{\Lambda
	\sqrt{1-\frac{2m}{R}}}\right)\right\}.
	\end{equation}
	This choice implies that, to zeroth order, we are dealing with a black hole space-time. This allows us to consider the limit in which the quantum scalar fields act as perturbations to the black hole space-time. 
	
	With this choice of $S^{(0)}$, it is straightforward to find a solution for $S^{(1)}$,
	\begin{equation}
	S^{(1)} = \int dr \, \frac{\phi_1+\phi_2}{2} \sqrt{\frac{C_F}{2}} \; + \; C^{(1)}[R,\Lambda],
	\end{equation}
	where $C^{(1)}$ is an arbitrary functional of the gravitational degrees of freedom. 
	\item First order terms:
	\begin{eqnarray}
&~&-\frac{1}{R} \left( \frac{\delta S^{(0)}}{\delta R}\frac{\delta S^{(1)}}{\delta
	\Lambda } + \frac{\delta S^{(0)}}{\delta \Lambda}\frac{\delta S^{(1)}}{\delta
	R }  \right)+ \frac{\Lambda }{2R^{2}}\left( 2 \frac{\delta S^{(0)}}{\delta \Lambda }\frac{\delta S^{(1)}}{\delta \Lambda }
\right)   \nonumber\\
	&~& + \sum_{a=1}^2 \left\{ \frac{1}{2\Lambda R^{2}}\left( \frac{\delta S^{(1)}}{\delta \phi_a }\frac{\delta S^{(2)}}{\delta \phi_a }\right) \right\} = 0.
	\end{eqnarray}
	
	The equation is linear in $S^{(2)}$. Except for ${\delta S^{(2)}}/{\delta \phi_a }$, all terms are known and they depend on $\phi_1+\phi_2$ only, so it is straightforward to solve for $S^{(2)}$. It is given by
	\begin{eqnarray}
	S^{(2)} &=& \frac{1}{2}\int dr \, (\phi_1+\phi_2) \left[ -\frac{1}{R} \left( \frac{\delta S^{(0)}}{\delta R}\frac{\delta S^{(1)}}{\delta \Lambda } + \frac{\delta S^{(0)}}{\delta \Lambda}\frac{\delta S^{(1)}}{\delta R }  \right) \right.   \nonumber\\
	&~& \left. + \frac{\Lambda }{2R^{2}}\left(2 \frac{\delta S^{(0)}}{\delta \Lambda }\frac{\delta S^{(1)}}{\delta \Lambda }
	\right)  \right]\frac{2\Lambda R^{2}}{\sqrt{C_F/2}} \;+\; C^{(2)}[R,\Lambda,\phi_1-\phi_2]\nonumber\\
	&=:& B^{(2)}[R,\Lambda,\phi_1+\phi_2] \;+\; C^{(2)}[R,\Lambda,\phi_1-\phi_2]
		\end{eqnarray}
	where $C^{(2)}$ is a quadratic but otherwise arbitrary functional of $(\phi_1-\phi_2)$. Notice that in general $S^{(2)} \neq S_1^{(2)}[R,\Lambda,\phi_1] + S_2^{(2)}[R,\Lambda,\phi_2]$, which implies \textit{entanglement of $\phi_1$ and $\phi_2$}, as per the discussion in section \ref{entanglement}. So already at the first order of the expansion we have in general entangled states\footnote{We are dealing here with a triple joint configuration space, corresponding to the two quantum fields ($\phi_1$ and $\phi_2$) plus the gravitational degrees of freedom (described by $\Lambda$ and $R$). But we are focusing here on the entanglement between the two quantum fields only. Thus we consider the entanglement between $\phi_1$ and $\phi_2$ relative to the classical gravity configuration, i.e., relative to a given $R$ and $\Lambda$.  In this case one is effectively dealing with $P$ and $S$ conditioned on $R$ and $\Lambda$.\blk}. Non-entangled states are only possible  for special choices of $C^{(2)}$.
	
	\item Second order terms:
	\begin{eqnarray}
&~&-\frac{1}{R} \left( \frac{\delta S^{(0)}}{\delta R}\frac{\delta S^{(2)}}{\delta
	\Lambda } + \frac{\delta S^{(0)}}{\delta \Lambda}\frac{\delta S^{(2)}}{\delta
	R } + 2 \frac{\delta S^{(1)}}{\delta \Lambda}\frac{\delta S^{(1)}}{\delta
	R }  \right) \nonumber\\
&~&+ \frac{\Lambda }{2R^{2}}\left( 2\frac{\delta S^{(0)}}{\delta \Lambda }\frac{\delta S^{(2)}}{\delta \Lambda } + \left( \frac{\delta S^{(1)}}{\delta \Lambda }
\right) ^{2}  \right)   \nonumber\\
	&~& + \sum_{a=1}^2 \left\{ \frac{1}{2\Lambda R^{2}}\left( 2 \frac{\delta S^{(1)}}{\delta \phi_a }\frac{\delta S^{(3)}}{\delta \phi_a } +\left( \frac{\delta S^{(2)}}{\delta \phi_a }
\right) ^{2} \right)  \right\}  \nonumber\\
	&~& + \sum_{a=1}^2 \left\{ \frac{R^2}{2\Lambda } (\phi_a')^2 + \frac{\Lambda R^2 m^2}{2} \phi_a^2 +\frac{1}{8 \Lambda R^2} \frac{1}{4} \left( \frac{\delta F^{(2)}}{\delta \phi_a} \right)^2 \right\}= 0.
	\end{eqnarray}
	
	The equation is linear in $S^{(3)}$ but much more complicated than the corresponding equation for $S^{(2)}$ that we solved above. We will not attempt an explicit solution here, but it is clear from the form of the equation that we will once more have in general solutions where $S^{(3)} \neq S_1^{(3)}[R,\Lambda,\phi_1] + S_2^{(3)}[R,\Lambda,\phi_2]$, implying also at this level of the expansion that there is  \textit{entanglement of $\phi_1$ and $\phi_2$}, as per the discussion in section \ref{entanglement}.
\end{enumerate}

A perturbative solution requires solving the continuity equation, Eq. (\ref{1142_04}), to the same order of the expansion of $F$ in powers of $\phi_a$ as we did for $S$. We will not carry out this step here, as the main purpose of the exercise, which is to show the existence of entangled states, has already been accomplished by looking at the solution of the EHJ equation. However, we nevertheless write down the equations that determine the terms $F^{(n)}$ up to order $n=2$, for completeness.
\begin{enumerate}
	\item Zeroth order terms (considering that $F^{(1)}=0$):
\begin{eqnarray} 
&~& \, \left\{
\frac{\delta F^{(0)}}{\delta R}\left( \frac{1}{R}\frac{\delta S^{(0)}}{\delta \Lambda }\right)
+\frac{\delta F^{(0)}}{\delta \Lambda }\left( \frac{1}{R}\frac{\delta S^{(0)}}{\delta R}
-P\frac{\Lambda }{R^{2}}\frac{\delta S^{(0)}}{\delta\Lambda }\right)\right.\nonumber\\
&~&\left.
+\frac{\delta }{\delta R}\left( \frac{1}{R}\frac{\delta S^{(0)}}{\delta \Lambda }\right)
+\frac{\delta }{\delta \Lambda }\left( \frac{1}{R}\frac{\delta S^{(0)}}{\delta R}
-\frac{\Lambda }{R^{2}}\frac{\delta S^{(0)}}{\delta\Lambda }\right)\right.\nonumber\\
&~&\left.- \sum_{a=1}^n  \left[ \frac{\delta }{\delta \phi_a }\left( \frac{1}{\Lambda R^{2}}
\frac{\delta S^{(2)}}{\delta \phi_a }\right) \right] \right\}=0.
\end{eqnarray}

This is an equation for $F^{(0)}$.

\item First order terms (considering that $F^{(1)}=0$):
\begin{eqnarray}
&~& \, \left\{
\frac{\delta F^{(0)}}{\delta R}\left( \frac{1}{R}\frac{\delta S^{(1)}}{\delta \Lambda }\right)
+\frac{\delta F^{(0)}}{\delta \Lambda }\left( \frac{1}{R}\frac{\delta S^{(1)}}{\delta R}
-P\frac{\Lambda }{R^{2}}\frac{\delta S^{(1)}}{\delta\Lambda }\right)\right.\nonumber\\
&~&\left.
+\frac{\delta }{\delta R}\left( \frac{1}{R}\frac{\delta S^{(1)}}{\delta \Lambda }\right)
+\frac{\delta }{\delta \Lambda }\left( \frac{1}{R}\frac{\delta S^{(1)}}{\delta R}
-\frac{\Lambda }{R^{2}}\frac{\delta S^{(1)}}{\delta\Lambda }\right)\right.\nonumber\\
&~&\left.- \sum_{a=1}^n  \left[  \frac{\delta F^{(2)} }{\delta \phi_a }\left( \frac{1}{\Lambda R^{2}}
\frac{\delta S^{(1)}}{\delta \phi_a }\right) +  \frac{\delta }{\delta \phi_a }\left( \frac{1}{\Lambda R^{2}}
\frac{\delta S^{(3)}}{\delta \phi_a }\right) \right] \right\}=0.
	\end{eqnarray}
	
This is an equation for $F^{(2)}$.	

\end{enumerate}

Therefore, it is possible in principle to solve for $F$ in powers of $\phi_a$ to the same order as for $S$. It is clear that the equations for $F$ are extremely complicated and other approaches are needed if one wants to carry out practical calculations. However, the analysis, even if incomplete in this sense, shows the existence of entangled states for the quantum fields via their interaction with a common classical gravitational field. 

\section{The emergence of time and entanglement}\label{EoT}

In the formalism used in the previous section, time does not play a role -- which is natural given that the our analysis is based on the Hamilton-Jacobi formulation of general relativity. 

However, one would like to discuss the time evolution of the quantized fields. This \textit{is} possible: since the gravitational field is treated classically, one may introduce a well defined \textit{gravitational time} and derive a time-dependent equation for the quantized fields. Thus one may look into the question of whether the coupling through the classical gravitational field can generate entanglement as the hybrid system evolves. This question is of particular relevance to the interpretation of the experimental proposals of Bose et al.~\cite{bose} and Marletto and Vedral~\cite{marl}, as discussed in the introduction. 

We will take a similar starting point to the previous section, but instead of a perturbation expansion we will make use of a suitable physical approximation. 
We write 
\begin{eqnarray}
S &=& S_A[R,\Lambda] + S_B[R,\Lambda,\phi_a],\label{SAB}\\
P &=& P_A[R,\Lambda] \, P_B[R,\Lambda,\phi_a],\label{PAB}
\end{eqnarray}
and choose $S_A$ and $P_A$ that are solutions of a classical space-time without the scalar fields (i.e., a vacuum space-time, in this case a black hole).  Thus our first step is to choose $S_A$ and $P_A$ that satisfy the Einstein-Hamilton-Jacobi and continuity equations for a black hole space-time. Then we look at an approximation that corresponds to a non-linear generalization of the functional Schr\"{o}dinger equation of quantum field theory on curved space-time for $\Psi := \sqrt{P_B} \exp\left(i S_B \right)$. The way this works will become clear as we carry out the steps in sections \ref{TEHJE} and \ref{TCE}.

\subsection{Gravitational time}

A solution $S[h_{ij}]$ of the Einstein-Hamilton-Jacobi equation, Eq. (\ref{Appendix-ehj}), is a functional of the metric $h_{ij}$ on a three-dimensional space-like hypersurface. It is always possible to use $S$ to reconstruct the four-dimensional \textit{space-time}. This requires introducing lapse and shift functions together with rate equations for $h_{ij}$ \cite{MTW73,K12}. 

Similar considerations apply to spherical gravity. In this case, a solution $S[\Lambda, R]$ of the Einstein-Hamilton-Jacobi equation, Eq. (\ref{1022-03c}), is a functional of $\Lambda$ and $R$. It is possible to reconstruct the two-dimensional space-time of spherical gravity via the rate equations \cite{C10_K94}
\begin{eqnarray}\label{1022_11}
\dot{R} &=&-\frac{N}{R}\frac{\delta S}{\delta \Lambda }+N_rR^{\prime },
\label{fldvelApp1} \\
\dot{\Lambda} &=&-\frac{N}{R}\frac{\delta S}{\delta R}+\frac{\Lambda }{R^{2}}
\frac{\delta S}{\delta \Lambda }+\left( \Lambda N_r\right) ^{\prime },
\end{eqnarray}
where $N$ is the lapse function and $N_r$ is the shift function in Eq.~(\ref{1022_01}). We will call the time that enters into the rate equations the \textit{gravitational time}. 

We will now introduce such a gravitational time, and use it to derive time-dependent equations for the evolution of the quantum fields (the procedure that we will follow is similar to one that has been applied in studies of the semi-classical approximation to quantum geometrodynamics \cite{K12,K93,KS91}). This first requires making a choice of lapse and shift functions, and so we set $N=1$ and $N_r=0$. Thus, 
\begin{eqnarray}\label{REGCC}
\dot{R} &=&-\frac{1}{R}\frac{\delta S_A}{\delta \Lambda },\\
\dot{\Lambda} &=&-\frac{1}{R}\frac{\delta S_A}{\delta R}+\frac{\Lambda }{R^{2}}
\frac{\delta S_A}{\delta \Lambda }.
\end{eqnarray}

\subsection{The EHJ equation}\label{TEHJE}

We choose $S_A$ to be the solution to the classical EHJ equation for a black hole, 
\begin{equation}
 -\frac{1}{R}\frac{\delta S_A}{\delta R}\frac{\delta S_A}{\delta
	\Lambda } + \frac{\Lambda }{2R^{2}}\left( \frac{\delta S_A}{\delta \Lambda }
\right) ^{2} +  \frac{RR^{\prime \prime }}{\Lambda }-\frac{RR^{\prime }\Lambda ^{\prime }}{\Lambda ^{2}}
+\frac{R^{\prime 2}}{2\Lambda }-\frac{\Lambda }{2} = 0,  
\end{equation}
so that $S_A$ is given by Eq. (\ref{1030_05}). Then, the remaining terms of the EHJ equation, Eq. (\ref{1142_04}), are given by
\begin{eqnarray}
&~&-\frac{1}{R} \left( \frac{\delta S_A}{\delta R}\frac{\delta S_B}{\delta
	\Lambda } + \frac{\delta S_A}{\delta \Lambda}\frac{\delta S_B}{\delta
	R }  \right) + \frac{\Lambda }{R^{2}}\left( \frac{\delta S_A}{\delta \Lambda }\frac{\delta S_B}{\delta \Lambda }
\right)   \nonumber\\
&~&-\frac{1}{R}\frac{\delta S_B}{\delta R}\frac{\delta S_B}{\delta
	\Lambda } + \frac{\Lambda }{2R^{2}}\left( \frac{\delta S_B}{\delta \Lambda }
\right) ^{2}   \nonumber\\
&~& + \sum_{a=1}^2 \left\{ \frac{1}{2\Lambda R^{2}}\left( \frac{\delta S_B}{\delta \phi_a }\right) ^{2} + \frac{R^2}{2\Lambda } (\phi_a')^2 + \frac{\Lambda R^2 m^2}{2} \phi_a^2 \right\} \nonumber\\
&~& - \sum_{a=1}^2 \left\{ \frac{1}{2\Lambda R^2} 
\; \frac{1}{\sqrt{P}}\frac{\delta ^{2}\sqrt{P}}{\delta \phi_a ^{2}}  \right\}\nonumber\\
&~& = 0.
\end{eqnarray}

Using the rate equations, we can put the remainding terms of the EHJ equation in the form
\begin{eqnarray}
&~& \frac{\delta S_B}{\delta \Lambda }\dot{\Lambda} + \frac{\delta S_B}{\delta
	R }\dot{R}  + \sum_{a=1}^2 \left\{ \frac{1}{2\Lambda R^{2}}\left( \frac{\delta S_B}{\delta \phi_a }\right) ^{2}   \right\}  \nonumber\\
&~& +  \sum_{a=1}^2 \left\{\frac{R^2}{2\Lambda } (\phi_a')^2 + \frac{\Lambda R^2 m^2}{2} \phi_a^2 -\frac{1}{2\Lambda R^2} 
\; \frac{1}{\sqrt{P}}\frac{\delta ^{2}\sqrt{P}}{\delta \phi_a ^{2}}  \right\} \nonumber\\
&~&-\frac{1}{R}\frac{\delta S_B}{\delta R}\frac{\delta S_B}{\delta
	\Lambda } + \frac{\Lambda }{2R^{2}}\left( \frac{\delta S_B}{\delta \Lambda }
\right) ^{2}   \nonumber\\
&~& = 0.
\end{eqnarray}
Finally, integrating with respect to $r$, we get
\begin{eqnarray}
\dot{S}_B &=& \int dr \left[ \sum_{a=1}^2 \left\{ \frac{1}{2\Lambda R^{2}}\left( \frac{\delta S_B}{\delta \phi_a }\right) ^{2}   \right\} \right. \nonumber\\
&+&  \left. \sum_{a=1}^2 \left\{\frac{R^2}{2\Lambda } (\phi_a')^2 + \frac{\Lambda R^2 m^2}{2} \phi_a^2 -\frac{1}{2\Lambda R^2} 
\; \frac{1}{\sqrt{P}}\frac{\delta ^{2}\sqrt{P}}{\delta \phi_a ^{2}}  \right\} \right.\nonumber\\
&-& \left. \frac{1}{R}\frac{\delta S_B}{\delta R}\frac{\delta S_B}{\delta
	\Lambda } + \frac{\Lambda }{2R^{2}}\left( \frac{\delta S_B}{\delta \Lambda }
\right) ^{2} \right],  
\end{eqnarray}
where we have used 
\begin{equation}
\int dr \, \frac{\delta S_B}{\delta \Lambda }\dot{\Lambda} + \frac{\delta S_B}{\delta R }\dot{R} = \dot{S}_B.
\end{equation}

\subsection{The continuity equation for weak scalar fields}\label{TCE}

In the case of the continuity equation, Eq. (\ref{1142_05}), the ansatz of Eqs. (\ref{SAB}) and (\ref{PAB}) leads to
\begin{eqnarray} 
&~& 
\frac{\delta }{\delta R}\left[ P_A P_B\frac{1}{R}\left(\frac{\delta S_A}{\delta \Lambda }+\frac{\delta S_B}{\delta \Lambda }\right)\right]
\nonumber\\
&~& 
+\frac{\delta }{\delta \Lambda }\left[ P_A P_B\frac{1}{R}\left(\frac{\delta S_A}{\delta R }+\frac{\delta S_B}{\delta R }\right)
-P_A P_B\frac{\Lambda }{R^{2}}\left(\frac{\delta S_A}{\delta R }+\frac{\delta S_B}{\delta R }\right)\right] \nonumber\\
&~& - \sum_{a=1}^2  \frac{\delta }{\delta \phi_a }\left(P_A P_B \frac{1}{\Lambda R^{2}}
\frac{\delta S_B}{\delta \phi_a }\right)  = 0.
\end{eqnarray}
where we used $\frac{\delta S_A}{\delta \phi_a }=0$.

We will make the assumption that 
\begin{equation} \label{approx}
\frac{\delta S_A}{\delta R }>>\frac{\delta S_B}{\delta R },~~~~~~~~~~\frac{\delta S_A}{\delta \Lambda }>>\frac{\delta S_B}{\delta \Lambda }.
\end{equation}
This condition amounts to assuming that the dependence on the gravitational variables is much weaker for $S_B$ than it is for $S_A$. Such a condition does not seem unreasonable: given that we are assuming that scalar fields can be treated as a perturbation of the black hole, one can expect $S_A$ to account for most of the gravitational field degrees of freedom and for $S_B$ to act as a correction term required by the presence of the quantized scalar field. Assuming these inequalities hold, we can neglect the terms that contain functional derivatives of $S_B$ with respect to $R$ and $\Lambda$ so that, in this approximation, the continuity equation becomes 
\begin{eqnarray} 
&~& 
\frac{\delta }{\delta R}\left( P_A P_B\frac{1}{R}\frac{\delta S_A}{\delta \Lambda }\right)
+\frac{\delta }{\delta \Lambda }\left( P_A P_B\frac{1}{R}\frac{\delta S_A}{\delta R }
-P_A P_B\frac{\Lambda }{R^{2}}\frac{\delta S_A}{\delta R }\right) \nonumber\\
&~& - \sum_{a=1}^2  \frac{\delta }{\delta \phi_a }\left( P_A P_B \frac{1}{\Lambda R^{2}}
\frac{\delta S_B}{\delta \phi_a }\right)  = 0,
\end{eqnarray}
which we write as
\begin{eqnarray} 
&~& P_B \left[
\frac{\delta }{\delta R}\left( P_A  \frac{1}{ R}\frac{\delta S_A}{\delta \Lambda }\right) +\frac{\delta  }{\delta \Lambda }\left(P_A   \frac{1}{R}\frac{\delta S_A}{\delta R} - P_A  \frac{\Lambda }{R^{2}}\frac{\delta S_A}{\delta R }\right)  \right] \nonumber\\
&~& + P_A \left[
\left(  \frac{1}{R}\frac{\delta S_A}{\delta \Lambda }\frac{\delta P_B}{\delta R}\right) +\left( \frac{1}{R}\frac{\delta S_A}{\delta R}\frac{\delta  P_B}{\delta \Lambda } - \frac{\Lambda }{R^{2}}\frac{\delta S_A}{\delta\Lambda }\frac{\delta  P_B}{\delta \Lambda }\right) \right. \nonumber\\
&~&\left.+ \sum_{a=1}^2  \frac{\delta }{\delta \phi_a }\left(  P_B\frac{1}{\Lambda R^{2}}
\frac{\delta S_B}{\delta \phi_a }\right) \right] \nonumber\\
&~& = 0.
\end{eqnarray}
We now choose $P_A$ so that $S_A$ and $P_A$ solve the continuity equation for the classical black hole. Then the term in square brackets on the first line is zero and the continuity equation reduces to
\begin{eqnarray} 
&~& \left(  \frac{1}{R}\frac{\delta S_A}{\delta \Lambda }\frac{\delta P_B}{\delta R}\right) +\left( \frac{1}{R}\frac{\delta S_A}{\delta R}\frac{\delta  P_B}{\delta \Lambda } - \frac{\Lambda }{R^{2}}\frac{\delta S_A}{\delta\Lambda }\frac{\delta  P_B}{\delta \Lambda }\right)  \nonumber\\
&~& + \sum_{a=1}^2  \frac{\delta }{\delta \phi_a }\left(  P_B\frac{1}{\Lambda R^{2}}
\frac{\delta S_B}{\delta \phi_a }\right)  = 0.
\end{eqnarray}

Using the rate equations and integrating over $r$, we can put the remaining terms in the continuity equation in the form
\begin{equation} 
\dot{P}_B = \int dr \, \sum_{a=1}^2  \frac{\delta }{\delta \phi_a }\left(  P_B\frac{1}{\Lambda R^{2}}
\frac{\delta S_B}{\delta \phi_a }\right),
\end{equation}
where we used
\begin{equation}
\int dr \, \frac{\delta P_B}{\delta \Lambda }\dot{\Lambda} + \frac{\delta P_B}{\delta R }\dot{R} = \dot{P}_B.
\end{equation}

\subsection{Wavefunctional representation}

If we now define the wavefunctional\footnote{Recalling that $S = S_A[R,\Lambda] + S_B[R,\Lambda,\phi_a]$ and $P = P_A[R,\Lambda] \, P_B[R,\Lambda,\phi_a]$ as per Eqs. (\ref{SAB}) and (\ref{PAB}), we have $P_B = P / P_A$ so that $\Psi$ corresponds in fact to a conditional wavefunctional \cite{hrbook}, conditioned on given values of $\Lambda$ and $R$.} 
\begin{equation}
\Psi[R,\Lambda,\phi_a;t) := \sqrt{P_B} \exp\left(i S_B \right),
\end{equation}
the modified EHJ and continuity equations can be written in terms of a \textit{non-linear functional Schr\"{o}dinger equation},
\begin{eqnarray}
i \hbar \dot{\Psi} &=& \hat{H}_f \Psi\nonumber\\
&=& \int dr \left[ \sum_{a=1}^2 \left\{ \frac{1}{2\Lambda R^{2}} \frac{\delta ^2}{\delta \phi_a ^2} + \frac{R^2}{2\Lambda } (\phi_a')^2 + \frac{\Lambda R^2 m^2}{2} \phi_a^2 \right\} + \Delta \right] \Psi, 
\label{nlse}
\end{eqnarray}
where the non-linear correction term $\Delta$ is given by
\begin{equation}
\Delta = - \frac{1}{R}\frac{\delta S_B}{\delta R}\frac{\delta S_B}{\delta
	\Lambda } + \frac{\Lambda }{2R^{2}}\left( \frac{\delta S_B}{\delta \Lambda }
\right) ^{2} ,
\end{equation}
($\Delta$ may of course be written in terms of $\Psi$ and $\bar{\Psi}$). $\Delta$ is a new ``correction'' term that distinguishes the time evolution as evaluated by quantum field theory in curved space-time (where this term is absent) from the time evolution in the hybrid theory.

We would like to consider the following question: suppose that initially the two quantum fields $\phi_1$ and $\phi_2$ are \textit{not} entangled. Can this non-linear time-dependent functional Schr\"{o}dinger equation lead to their entanglement? 

The crucial point here is that there is no reason to believe that the term $\Delta$ will preserve non-entanglement of states, as it is quadratic in the functional derivatives of $S_B[R,\Lambda,\phi_1,\phi_2]$ with respect to $R$ and $\Lambda$. 

One can argue as follows. Consider calculating the time evolution of the wavefunctional $\Psi$ after an infinitesimally small time interval $\delta t$. If the initial state is not entangled so that
\begin{equation}
S_B[R,\Lambda,\phi_1,\phi_2;t=0)=S_B^1[R,\Lambda,\phi_1;t=0)+S_B^2[R,\Lambda,\phi_2;t=0),
\end{equation}
the initial $\Delta$ will have in general mixed terms in $\phi_1$ and $\phi_2$ which can lead to entanglement, so that one would expect at time $\delta t$ that
\begin{equation}
S_B[R,\Lambda,\phi_1,\phi_2;t=\delta t) \neq S_B^1[R,\Lambda,\phi_1;t=\delta t)+S_B^2[R,\Lambda,\phi_2;t=\delta t).
\end{equation}

This suggests that the interaction of $\phi_1$ and $\phi_2$ via a common gravitational field can cause in general entanglement.

\section{Discussion}\label{Discussion}

Our main result is that entanglement between quantum fields may be generated via a classical gravitational interaction (section \ref{EoT}). This result is based on the configuration ensemble formalism (which is able to describe the coupling of quantum and classical systems more generally), with explicit calculations made for the case of black-hole spacetimes in spherical gravity under a weak-field approximation as per Eq.~(\ref{approx}). The effective evolution equation for the quantum fields, Eq.(\ref{nlse}), is defined with respect to a gravitational time.

The above result strongly supports the arguments made in~\cite{HR2018}, that observation of entanglement {\it per se}, in the experiments proposed by Bose {\it et al.} and by Marletto and Vedral \cite{bose, marl}, does not necessarily imply that gravity is nonclassical in nature. Such an observation can only rule out some classical models of gravity, such as Koopmanian and mean-field models~\cite{HR2018}, but not all. In particular, entanglement appears to be ubiquitous in the configuration-ensemble model, as exemplified by the spherical gravity solutions in section~\ref{EntQFBH} and the approximate evolution equation in section~\ref{EoT}.

It should be noted that whereas the  configuration-ensemble model used for our calculations took the standard canonical formalism for classical gravity as its starting point (see section \ref{vacgrav}), one could also carry out similar calculations based on an analogous model starting from the alternative canonical formalism due to Ashketar~\cite{ashketar}. Generation of entanglement is similarly expected for this case.

In future work, it will be important to consider the Newtonian limit of the configuration-ensemble model for quantum fields coupled to classical spacetime, in which the fields reduce to localised single-particle excitations and the gravitational interaction becomes Newtonian. This will allow quantitative predictions to be made in the context of the recent experimental proposals, and compared to those made via Newtonian~\cite{bose} and semiclassical~\cite{CR2018} models of gravity.

It will also be of interest to develop the configuration-ensemble model further, to study for example Hawking radiation and black hole evaporation using some of the techniques explored in sections~\ref{EntQFBH} and \ref{EoT} (which would allow one to extend results previously obtained for a classical CGHS black hole coupled to a quantized scalar field \cite{hrbook}), and to examine hybrid cosmological models. Such calculations, based on the configuration-ensemble model, could provide hints for developing a full quantum theory of gravity as well as suggesting experimental tests for quantum gravity.

\end{document}